\documentclass[twocolumn,showpacs,fleqn,nobibnotes]{revtex4}
\usepackage{amsmath}
\usepackage{graphicx}
\usepackage{float}
\usepackage{subfigure}

\def\lsim{\raise0.3ex\hbox{$<$\kern-0.75em\raise-1.1ex\hbox{$\sim$}}}
\def\gsim{\raise0.3ex\hbox{$>$\kern-0.75em\raise-1.1ex\hbox{$\sim$}}}

\begin{document}

\title{Photoproduction of Upsilon states in ultraperipheral collisions at the CERN Large Hadron Collider within the color dipole approach}
\author{M.B. Gay Ducati, F. Kopp, M.V.T. Machado and S. Martins}

\affiliation{High Energy Physics Phenomenology Group, GFPAE  IF-UFRGS \\
Caixa Postal 15051, CEP 91501-970, Porto Alegre, Rio Grande do Sul, Brazil}

\begin{abstract}
The exclusive photoproduction of Upsilon state $\Upsilon(1S)$ and its radially excited states $\Upsilon(2S,3S)$ is investigated  in the context of ultra-peripheral collisions at the LHC energies. Predictions are presented for their production in proton-proton, proton-nucleus and nucleus-nucleus collision at the energies available at the LHC run 2. The rapidity and transverse momentum distributions are shown, and the robustness of the model is tested against the experimental results considering $\psi(1S,2S)$  and $\Upsilon(1S)$ states. The theoretical framework considered in the analysis is the light-cone color dipole formalism, which includes consistently parton saturation effects and nuclear shadowing corrections.

\end{abstract}

\maketitle

\section{Introduction}

 The exclusive photoproduction of heavy vector mesons is a kind of diffractive process where, besides a soft scale characterized by the hadron size, there is clearly a hard scale (mesons mass $m_V$) that allows to analyse the reaction from the perturbative QCD point of view. This advantage creates ways to investigate the pomeron exchange which could lead to a better understanding of this object in terms of QCD. Other advantage in studying this process occurs in ultra-peripheral collisions \cite{upcs}, where the impact parameter is larger than the sum of the radius of the interacting hadrons. In this case, the exclusive photoproduction dominates the process through the emission of virtual photons which interact with the target featuring the photon-target cross section. The photon-target interaction amplitude, when considering the light-cone dipole formalism \cite{nik}, can be written as a convolution between the photon-meson wave functions overlap and the elementary dipole-target cross section \cite{Nemchik}. The exclusive quarkonium photoproduction has been investigated both
experimentally and theoretically in recent years. It allows to test
perturbative Quantum Chromodynamics as the  masses of these heavy
mesons give a perturbative scale for the problem even in the
photoproduction limit.  An important feature of these exclusive
processes at the high energy regime is the possibility to investigate
the hard pQCD Pomeron exchange. At present energy regime at LHC photons can
be considered as color dipoles in the mixed light cone representation,
where their transverse size can be considered frozen during the
interaction \cite{nik}. The corresponding scattering process is characterized by the
color dipole cross section describing the interaction of those color
dipoles with the nucleon or nucleus target. The referred approach is quite intuitive
and brings information on the dynamics beyond the leading
logarithmic QCD approach. The dynamics related to the meson formation is given by their wave-functions and to compute predictions for their excited states  is a reasonably task \cite{Nemchik}.

In this work, we investigate the exclusive production of
$\Upsilon (1S)$ and its radially excited states $\Upsilon (2S)$ and $\Upsilon (3S)$
in proton-proton, proton-nucleus and nucleus-nucleus collisions in the LHC energy interval. In previous works \cite{GGM1,GGM2} some of us have considered coherent photoproduction of $J/\psi$ and $\psi(2S)$ states at various energies in $pp$ and PbPb (lead-lead) collisions at the LHC. Those calculations were carried out in the theoretical framework of the color light-cone dipole formalism \cite{nik}. It was shown that the corresponding predictions describe nicely the experimental results from ALICE \cite{ALICE1,ALICE2} and LHCb \cite{LHCb1,LHCb2} collaborations on the charmonia production. In the dipole framework,  the heavy quark-antiquark fluctuation of the incoming
quasi-real photons interacts with the target via the
dipole cross section, and the result is projected into the wave-function
of the observed meson state. The photoproduction also gives us ways to investigate the transition between the linear dynamics, governed by DGLAP (Gribov-Lipatov-Altarelli-Parisi) and BFKL (Balitskii-Fadin-Kuraev-Lipatov) evolution equations, and the non-linear dynamics where the physical process of the partonic recombination, e.g. $gg\rightarrow g$, becomes important. In the energies available at LHC, the transition of the regime
described by the linear dynamics of emissions chain to a new regime,
where the physical process of recombination of partons becomes
important, is expected. It is characterized by the limitation on the
maximum phase-space parton density that can be reached in the hadron
wave-function, the  so-called parton saturation phenomenon  \cite{hdqcd,Gayreport}. The
transition is set by the saturation scale $Q_{\mathrm{sat}}$, which is enhanced in the nuclear case. Therefore, the color dipole approach will include both the parton saturation effects in photon-proton interaction as nuclear shadowing effects in photon-nucleus process. As examples of the values of Bjorken $x$ reached at LHC (at mid-rapidity) one has $x\simeq 10^{-4}$ for $pp$,  $x\simeq 10^{-3}$ for $pA$ and $x\simeq 2\times 10^{-3}$ for $AA$ collisions. Those values are even smaller for very forward rapidities, e.g. for $y=4$ the $x$-value diminishes by a factor one hundred.

The paper is organized as follows. In the next section we summarize the main theoretical information to compute the rapidity and transverse momentum distributions of $\Upsilon (1S,2S,3S)$ states in $pp$, $pA$ and $AA$ collisions for present and future runs at the LHC. We directly compare the results to the experimental data measured by the  LHCb Collaboration \cite{UpsilonLHCb} for $\Upsilon (1S)$ in $pp$ collisions. In the section \ref{discussions} we present the numerical calculations, discuss the main theoretical uncertainties and a comparison with other approaches is done. In particular, we contrast the main results  against the predictions available using the STARlight Monte Carlo \cite{STARlight,KN,dalsin1}. In the last section we show the main conclusions.

\section{Theoretical Framework}

At high energies, the total cross section for photoproduction of a vector meson $V$ in ultra-peripheral $pp$, $pA$ or $AA$ collisions  can be written in a factorized way as follows \cite{upcs,KN}
\begin{eqnarray}
\sigma_{V} = \int d\omega\frac{dN_{\gamma}}{d\omega}\,\sigma^{p(A)}_{\gamma}(\omega),\label{primeira} 
\end{eqnarray}
where the first term, $dN(\omega)/d\omega$, represents the virtual photons flux and can be calculated from the Weizs\"{a}cker-Williams method \cite{upcs}. It characterizes the photon energy ($\omega$)  distribution emitted by the hadrons. Moreover, the  photoproduction cross section, $\sigma^{p(A)}_{\gamma}(\omega)$, quantifies the cross section for the $\gamma + p(A)\rightarrow p(A) + V$ process.

 For highly energetic protons as those produced by the  LHC beams the photon flux can be approximated by \cite{upcs,KN}
\begin{eqnarray}
\frac{dN_{\gamma}^{p}}{d\omega} & = & \frac{\alpha_{em}}{2\pi\omega}\left[1+\left(1-\frac{2\omega}{\sqrt{s}}\right)^2\right]
\\
& \times & \left[\text{ln }\Omega-\frac{11}{6}+\frac{3}{\Omega}-\frac{3}{2\Omega^2}+\frac{1}{3\Omega^3}\right]
\end{eqnarray}
where $\Omega  = 1 + 0.71\text{ GeV}^2/Q^2_{min}$ e $Q^2_{min}=\left(\omega/\gamma_L\right)^2$, with $\gamma_L=\sqrt{s}/\left(2m_p\right)$. On the other hand, for nucleus having charge $Z$, the photon flux is approximately given by \cite{upcs,KN} 
\begin{eqnarray}
\frac{dN^{A}_{\gamma}}{d\omega} & = & \frac{2Z^2\alpha_{em}}{\pi}
\left[\xi K_0(\xi)K_1(\xi)-\frac{\xi^2}{2}\left(K_1^2(\xi)-K^2_0(\xi)\right)\right],\nonumber \\
& &\!\! 
\end{eqnarray}
where $\xi=\omega(R_p+R_A)/\gamma_L$ for $pA$ collisions and $\xi=\omega(2R_A)/\gamma_L$ for $AA$ collisions.

In the present work, we consider the color dipole approach for modeling the photon-target interaction. Therefore, the cross section $ \sigma^{p(A)}_{\gamma}$ appearing in Eq. ($\ref{primeira}$) will be computed from the following scattering amplitude in the photoproduction limit ($Q^2=0$) \cite{Nemchik},
\begin{eqnarray}
A^{\gamma^*p(A)\rightarrow Vp(A)}(x,\Delta)  & =&  \int d^2{\bf r}\int^1_0\frac{dz}{4\pi}\rho_V(z,r)A_{q\bar{q}}(x,r,\Delta),\nonumber \\
\!\!
\label{segunda}
\end{eqnarray}
where the quantity $\rho_V = \left(\Psi^*_V\Psi\right)_{T}$ represents the overlap of the photon-meson wave functions and $A_{q\bar{q}}$ is the dipole-target scattering amplitude (assumed to be  imaginary). Accordingly, the usual kinematical variables are as follows: $x$ is the Bjorken variable and the squared momentum transfer in hadron vertex is $t = -\Delta^2$. The variables $z$ and $r$ are the longitudinal momentum  fraction carried by the quark and the transverse color dipole size, respectively.

From Eq. ($\ref{segunda}$), we will consider the simplification that the $\Delta$ dependence on the amplitude is exponential, $A_{q\bar{q}}\propto e^{-B_V\Delta^2/2}$. Moreover, one includes the needed corrections related to the real part of amplitude, $\beta = Re\, A/Im\, A$, and off-diagonal momenta of exchanged gluons, $R_g$. Both quantities depend on the effective energy power behavior, $\lambda_{eff}$, of scattering amplitude. Thus, the cross section for  the exclusive photoproduction for $pp$ collisions reads as \cite{GGM1,GGM2},
\begin{eqnarray}  
\sigma\left(\gamma p\rightarrow Vp\right)(\omega )=\frac{R_g^2}{16\pi B_V}\left|A_p(x,\Delta=0)\right|^2 \left(1+\beta^2\right).
\end{eqnarray}

In the above equation, $B_V$ is the slope parameter which characterizes the size of interaction region. Along the calculation performed here, it was considered the energy dependence from the Regge phenomenology \cite{JPG42-105001} to describe the slope parameter,
\begin{eqnarray}
B_V\left(W_{\gamma p}\right)=b^V_{el}+2\alpha'\text{log}\left(\frac{W^2_{\gamma p}}{W^2_0}\right) ,
\end{eqnarray}
with $\alpha'=0.164$ GeV$^{-2}$, $W_{0}=95$ GeV, $b_{el}^{\Upsilon_{(1S)}}=3.68$ GeV$^{-2}$, $b_{el}^{\Upsilon_{(2S)}}=3.61$ GeV$^{-2}$ and $b_{el}^{\Upsilon_{(3S)}}=3.57$ GeV$^{-2}$.

In the case of nuclear targets, the coherent photonuclear cross section will be calculated using the following expression \cite{STARlight,KN},
\begin{eqnarray}
\sigma\left(\gamma A\rightarrow VA\right)(\omega) & = & R_g^2\frac{\left|A_{nuc}(x,\Delta=0)\right|^2}{16\pi}\left(1+\beta^2\right) \nonumber
\\
& \times & \int^{\infty}_{t_{min}} |F(t)|^2 \,dt,
\end{eqnarray}
where $t_{min}=\left(m^2_V\right/2\omega)^2$ and the nuclear form factor is given by
\begin{eqnarray}      
F (q)= \frac{4\pi\rho_0}{A\,q^3}\left[\sin \left(qR_A\right)-qR_A\cos \left(qR_A\right)\right]\left[\frac{1}{1+a^2q^2}\right],
\end{eqnarray} 
where $q=\sqrt{|t|}$, with  $\rho_0=0.16\text{ fm}^{-3}$ and $a=0.7$ fm \cite{KN}. 

\begin{figure}[t]
\begin{center}
\includegraphics[scale=0.28]{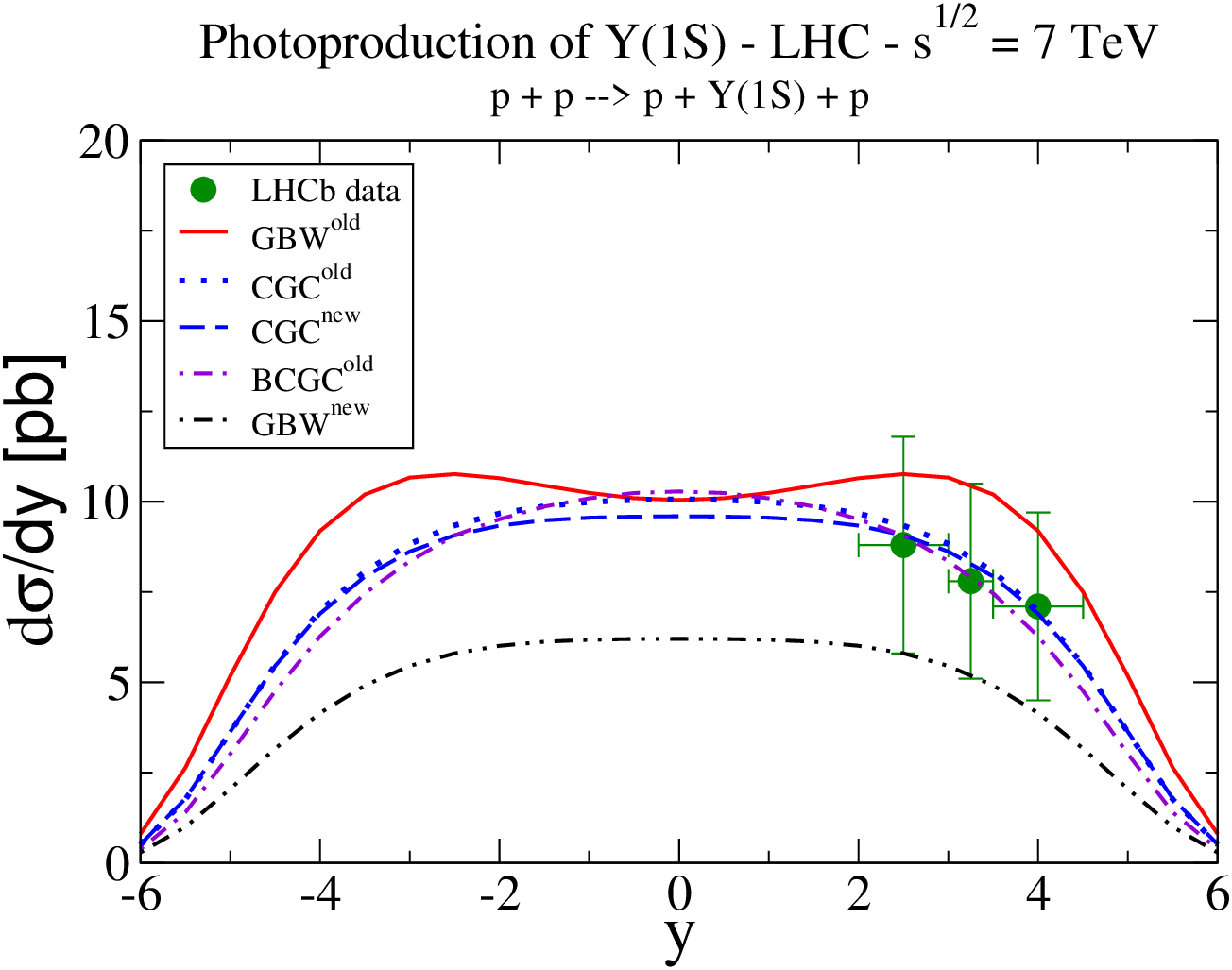} 
\end{center}
\caption{Rapidity distribution for the $\Upsilon(1S)$ state in $pp$ collisions at $\sqrt{s}=7$ TeV. Data from LHCb Collaboration\cite{UpsilonLHCb}}.
\label{dsigdyppY1S}
\end{figure}

Setting the vector meson wave-function (appearing in equation (\ref{segunda})) as a quark-antiquark state having spin and polarization structure similar to the photon, the corresponding overlap photon-meson wave-function can be written as \cite{wfbg}
\begin{eqnarray}
\rho_V (r,z) & = & \hat e_fe\frac{N_c}{\pi z(1-z)}\left\{m^2_fK_0(\varepsilon r)\phi_T(r,z)\right. \nonumber
\\
& - & \left.\left[z^2+(1-z)^2\right]\varepsilon K_1(\varepsilon r)\partial_r\phi_T(r,z)\right\}, \nonumber
\\
\end{eqnarray}
where the effective charge, $\text{\^e}_f=-1/3$, for the $\Upsilon $ states. In contrast  to the photon wave-functions which can be completely computed from perturbation theory \cite{nik}, the meson wave-function carries a phenomenological input embedded in functions $\phi_{T,L}$. In our calculations, we use the Boosted-Gaussian model \cite{wfbg} because it can be applied in a systematic way for excited states. The corresponding function is given by \cite{Sanda2},
\begin{eqnarray}
\phi_{nS}(r,z) = \left[\sum^{n-1}_{k=0}\alpha_{nS,k}R^2_{nS}\hat{D}^{2k}(r,z)\right]G_{nS}(r,z), \nonumber
\\
\end{eqnarray}
with $\alpha_{nS,0}=1$. The operator $\hat{D}^{2}(r,z)$ is defined by
\begin{eqnarray}
\hat{D}^{2}(r,z) = \frac{m_f^2-\nabla^2_r}{4z(1-z)}-m_f^2,
\end{eqnarray}
where $\nabla^2_r=\frac{1}{r}\partial_r+\partial^2_r$, and acts on the Gaussian function 
\begin{eqnarray}
 G_{nS}(r,z) & = & \mathcal{N}_{nS}\,z(1-z)\,\exp\left(-\frac{m^2_f\mathcal{R}^2_{nS}}{8z(1-z)} \right. \nonumber \\
& - & \left. \frac{2z(1-z)r^2}{\mathcal{R}^2_{nS}}+\frac{m^2_f\mathcal{R}^2_{nS}}{2}\right).
\end{eqnarray}

\begin{figure*}[t]
\includegraphics[scale=0.25]{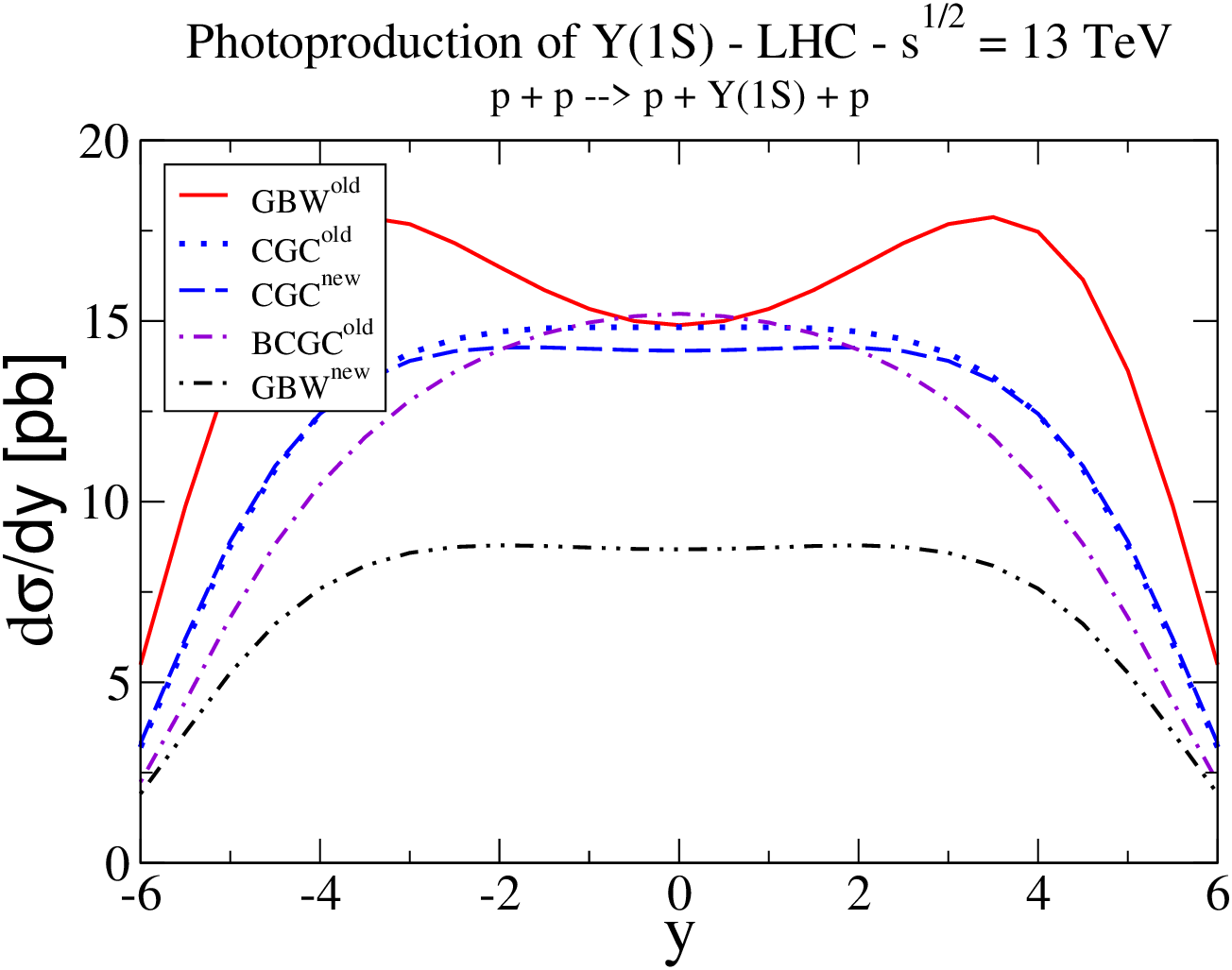} 
\includegraphics[scale=0.25]{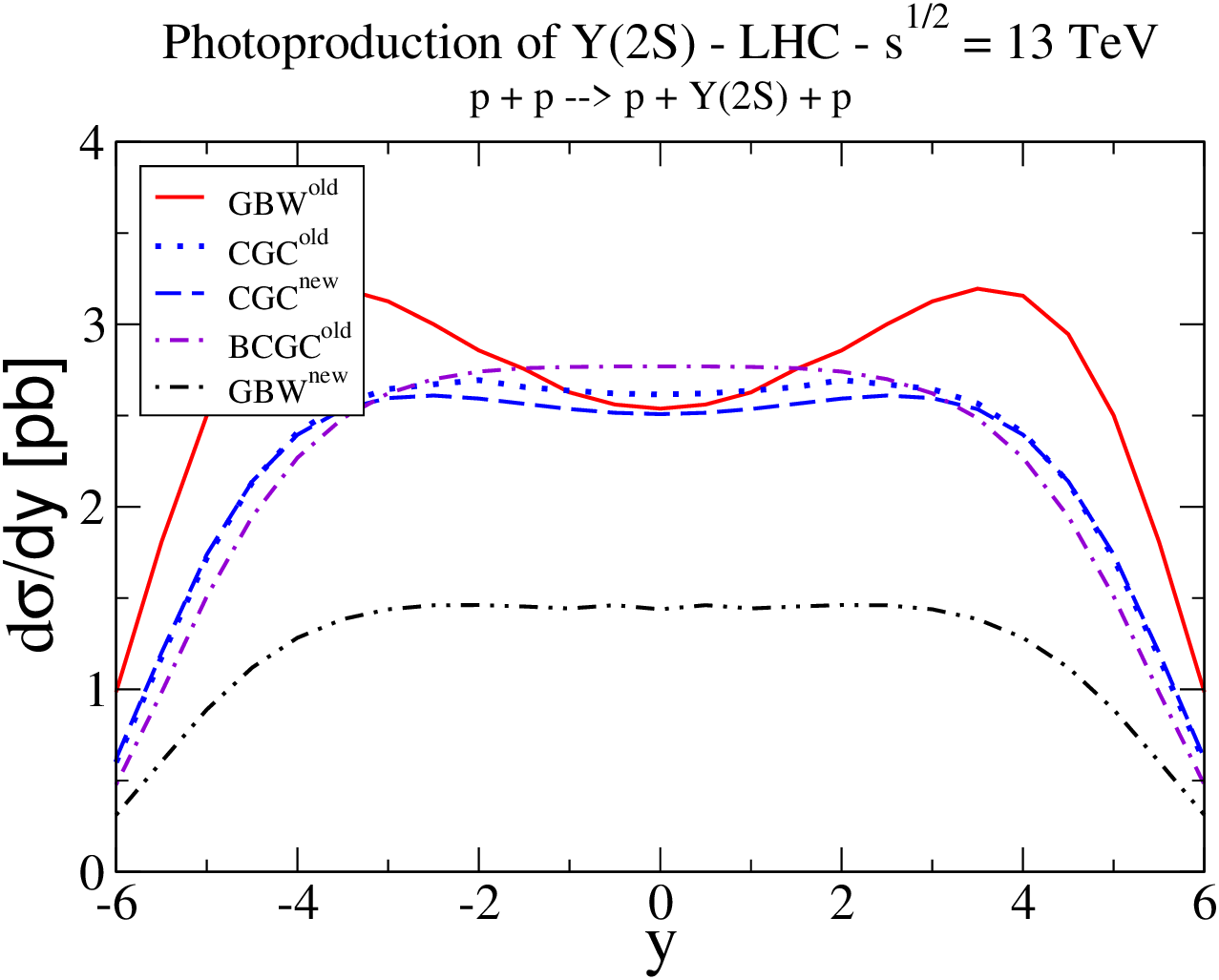}
\includegraphics[scale=0.25]{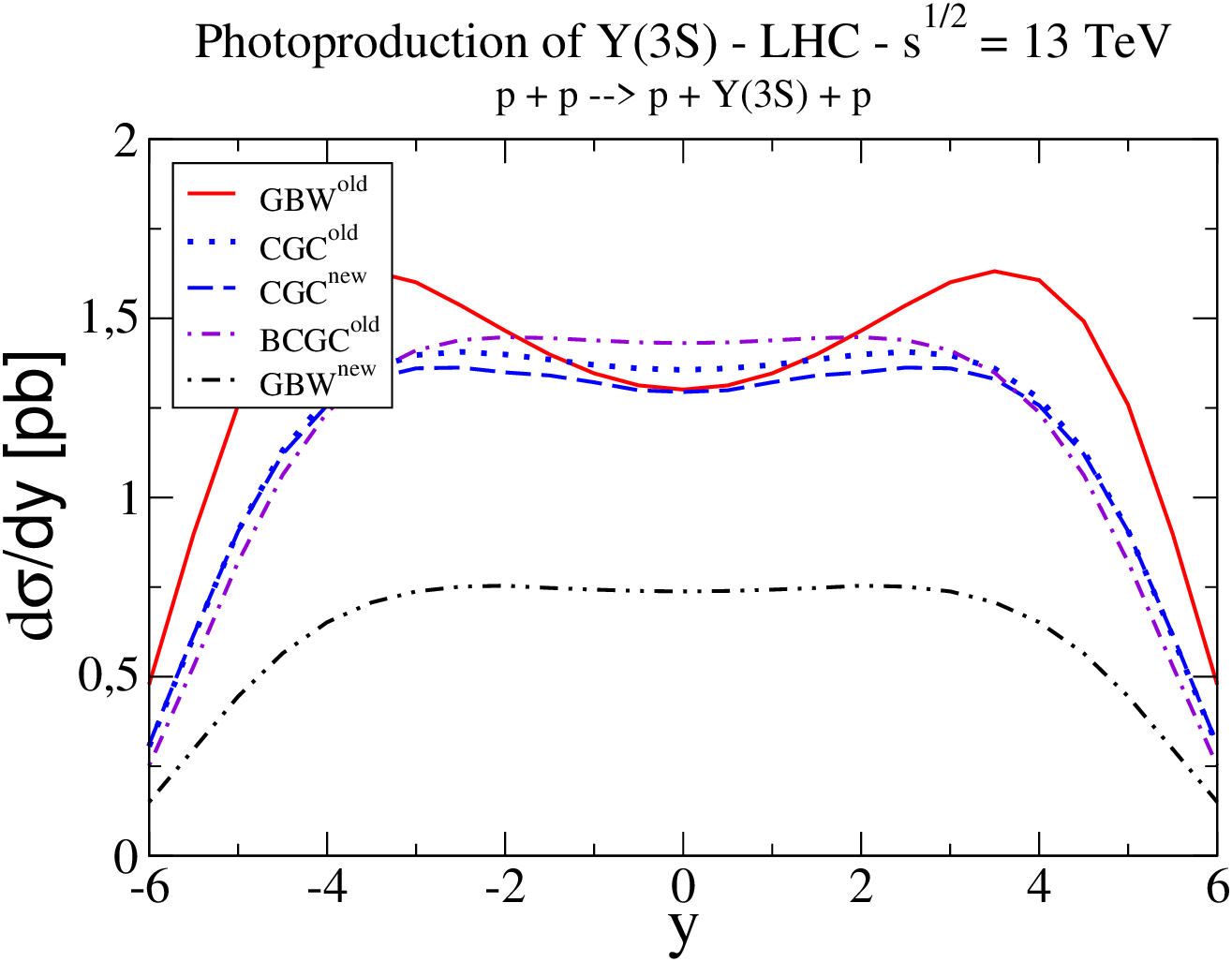} 
\caption{Rapidity distribution for the vector meson states  $\Upsilon(1S)$ (left panel), $\Upsilon(2s)$ (central panel) and $\Upsilon(3S)$ (right panel)  in $pp$ collisions in the LHC Run II at $\sqrt{s}=13$ TeV. Same notation as the previous figure.}
\label{Fig2}
\end{figure*}

The parameters $\mathcal{R}^2_{nS}$ and $\mathcal{N}_{nS}$ are determined by the normalization conditions and by the decay widths into dileptons \cite{Sanda2}. In addition, the parameters $\alpha_{nS,k}$ (for $k > 0$) are fixed by requiring that the wave-functions of different states be orthogonal to each other. For the boosted Gaussian, these parameters were calculated in \cite{Sanda2} and they are summarized in Table \ref{parametrosRN}. We call attention for the fact that the remaining literature, in general, uses only information on $\Upsilon (1S)$ state and constrain the excited state cross section using the measured values available at lower energies as provided by DESY-HERA experiment. 

\begin{table}[t]
\begin{tabular}{ccccccc}
\hline\hline
$\text{Meson}$ & $m_f/GeV$  & $M_V/GeV$  & $\mathcal{N}_T$ & $\mathcal{R}_{nS}^2/GeV^{-2}$  & $\alpha_{1S}$ & $\alpha_{2S}$\\ 
\hline
$\Upsilon(1S)$ & 4.2 & 9.46   & 0.481 & 0.567 &    -   &   -   \\
$\Upsilon(2S)$ & 4.2 & 10.023 & 0.624 & 0.831 & -0.555 &   -   \\
$\Upsilon(3S)$ & 4.2 & 10.355 & 0.668 & 1.028 & -1.219 & 0.217 \\
\hline\hline
\end{tabular}
\caption{Parameters  for Boosted Gaussian wave-function for $\Upsilon(1S)$, $\Upsilon(2S)$ and $\Upsilon(3S)$ states.}
\label{parametrosRN}
\end{table}

Other important component in Eq. (\ref{segunda}) is the dipole scattering amplitude. We use the following phenomenological models in our analyses: GBW (Golec-Biernat and Wusthoff) \cite{PRD59-014017}, GBW-KSX (Golec-Biernat, Wusthoff, Kozlov, Shoshi and Xian) \cite{JHEP0710-020}, CGC(Color Glass Condensate) \cite{PLB590-199} and BCGC(CGC with impact parameter dependence)\cite{KMW}. The dipole amplitude is related to the color dipole cross section in the form,
\begin{eqnarray}
\sigma_{q\bar q}(x,r) = 2\,\int d^2b \,A_{q\bar{q}}(x,r,b),
\end{eqnarray}
bearing in mind that $b$ and $\Delta$ are Fourier conjugates variables.

The GBW model is defined by the ansatz
\begin{eqnarray}
\sigma^{GBW}_{q\bar q}(x,r)=\sigma_0\left(1-e^{-r^2Q^2_s(x)/4}\right)\label{gbw},
\end{eqnarray}
where $\sigma_0=2\pi R^2$ is a constant, and $Q_s^2(x)=(x_0/x)^{\lambda}$ GeV$^2$ denotes the saturation scale, where the partonic recombination effects become important. A further correction has been introduced in the above model, called GBW-KSX  model\cite{JHEP0710-020}, including gluon number fluctuation effects. We also consider the CGC model \cite{PLB590-199}, based in the Color Glass Condensate framework, in which gluon saturation effects are incorporated via an approximate solution of the Balitsky-Kovchegov equation \cite{NPB463-99,PRD60-034008,PRD61-074018}. In this trend, two cases were considered: CGC-old parameterization \cite{PLB655-32}, which considers the previous DESY-HERA data, and CGC-new \cite{PRD88-074016} which considers more recent data from ZEUS and H1 combined
results for inclusive deep inelastic scattering. All the parameters associated to the referred models for the dipole amplitude are shown in Table \ref{tabparameters}. The expression for the CGC model is given by,
\begin{equation} \label{eq:cgc}
\sigma_{q\bar{q}}^{CGC}(x,r) = \sigma_0\begin{cases}
\mathcal{N}_0\left(\frac{rQ_s}{2}\right)^{\gamma_{eff}(x,r)} & :\quad rQ_s\le 2\\
1-\mathrm{e}^{-A\ln^2(BrQ_s)} & :\quad rQ_s>2
\end{cases},
\end{equation}
where $\gamma_{eff}(x,r)=2\left(\gamma_s+(1/\kappa\lambda Y)ln(2/rQ_s)\right)$ is the effective anomalous dimension, $Y=ln(1/x)$ and one has the constant $\kappa=9.9$. The saturation scale takes the same form as in GBW model.
Finally, we have tested also the impact parameter CGC  model (named b-CGC model). The expression for the b-CGC model is given by, 
\begin{equation} \label{eq:bcgc}
\sigma_{q\bar{q}}^{BCGC}(x,r)=2\int d^2b \,\begin{cases}
\mathcal{N}_0\left(\frac{rQ_s}{2}\right)^{\gamma_{eff}(x,r)} & :\quad rQ_s\le 2\\
1-\mathrm{e}^{-A\ln^2(BrQ_s)} & :\quad rQ_s>2
\end{cases},
\end{equation}
where the parameter $Q_s$ depends on the impact parameter:
\begin{equation} \label{eq:bcgc1}
Q_s\equiv Q_s(x,b)=\left(\frac{x_0}{x}\right)^{\frac{\lambda}{2}}\;\left[\exp\left(-\frac{b^2}{2B_{\rm CGC}}\right)\right]^{\frac{1}{2\gamma_s}},
\end{equation}

\begin{table}[t]
\begin{tabular}{c|cccc}
\hline\hline
& $\sigma_0$ (mb) & $x_0$ $\left(10^{-4}\right)$ & $\lambda$ & $\gamma_s$\\ 
\hline\hline                              
$GBW$        & 29.12 & 0.41   & 0.277  &   -    \\
$GBW^{ksx}$    & 31.85 & 0.0546 & 0.225  &   -    \\
$CGC_{old}$  & 27.33 & 0.1632 & 0.2197 & 0.7376 \\
$CGC_{new}$  & 21.85 & 0.6266 & 0.2319 & 0.762  \\
$BCGC_{old}$ &   -   & 0.0184   & 0.119  & 0.46 \\
\hline\hline
\end{tabular}
\caption{Parameters and charm quark mass associated to the distinct model for the dipole amplitude (see text). For the bottom mass one considers the fixed value $m_b=4.5$ GeV.}
\label{tabparameters}
\end{table}
\begin{flushleft}
where $B_{CGC}=7.5 \ GeV^{-2}$. \\
In the next section, we provide predictions to the future (current) runs of LHC for the rapidity and transverse momentum distributions of the $\Upsilon (1S,2S,3S)$ states in ultra-peripheral collisions.
\end{flushleft}

\section{Results and discussions}
\label{discussions}
\begin{figure*}[t]
\includegraphics[scale=0.25]{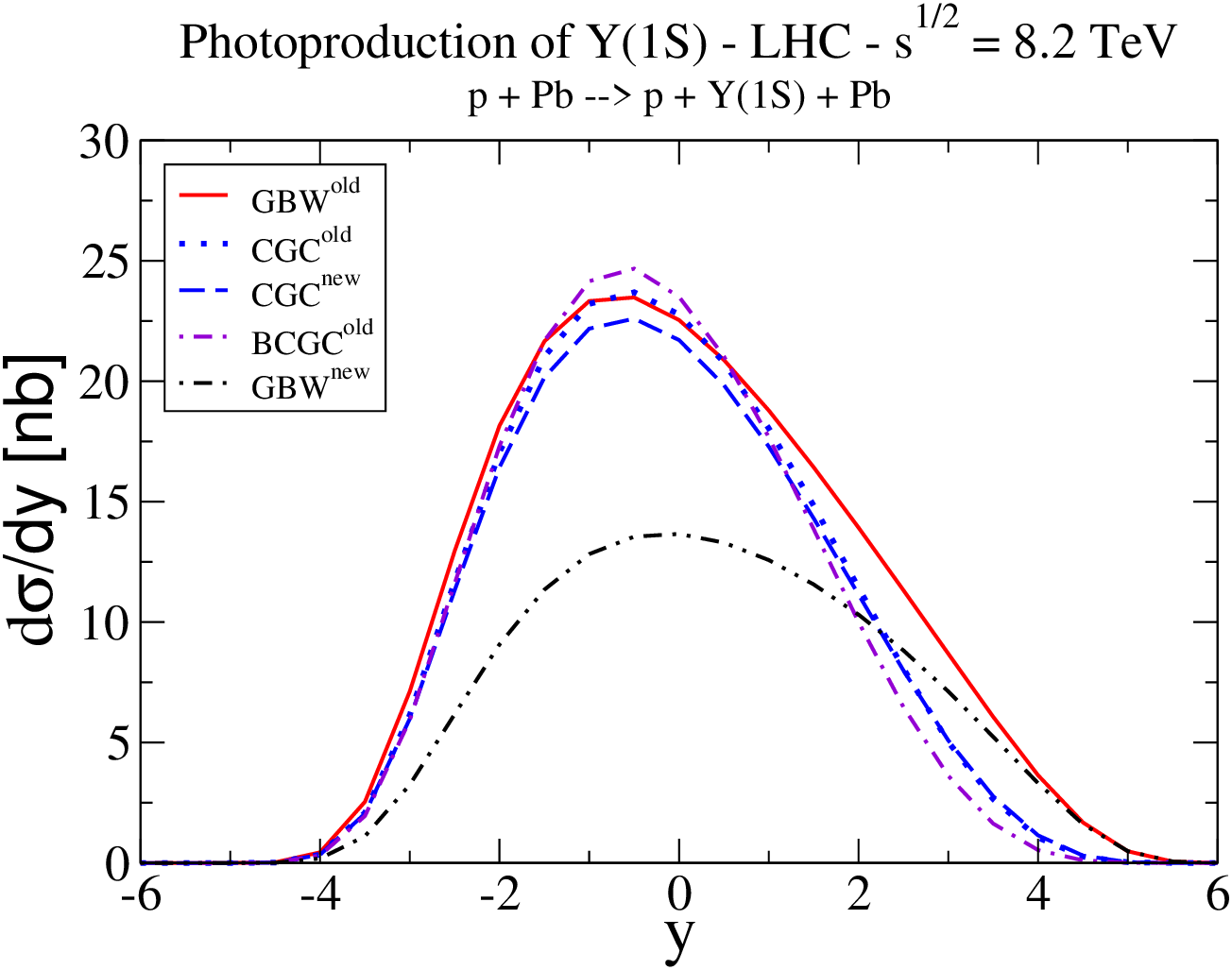} 
\includegraphics[scale=0.25]{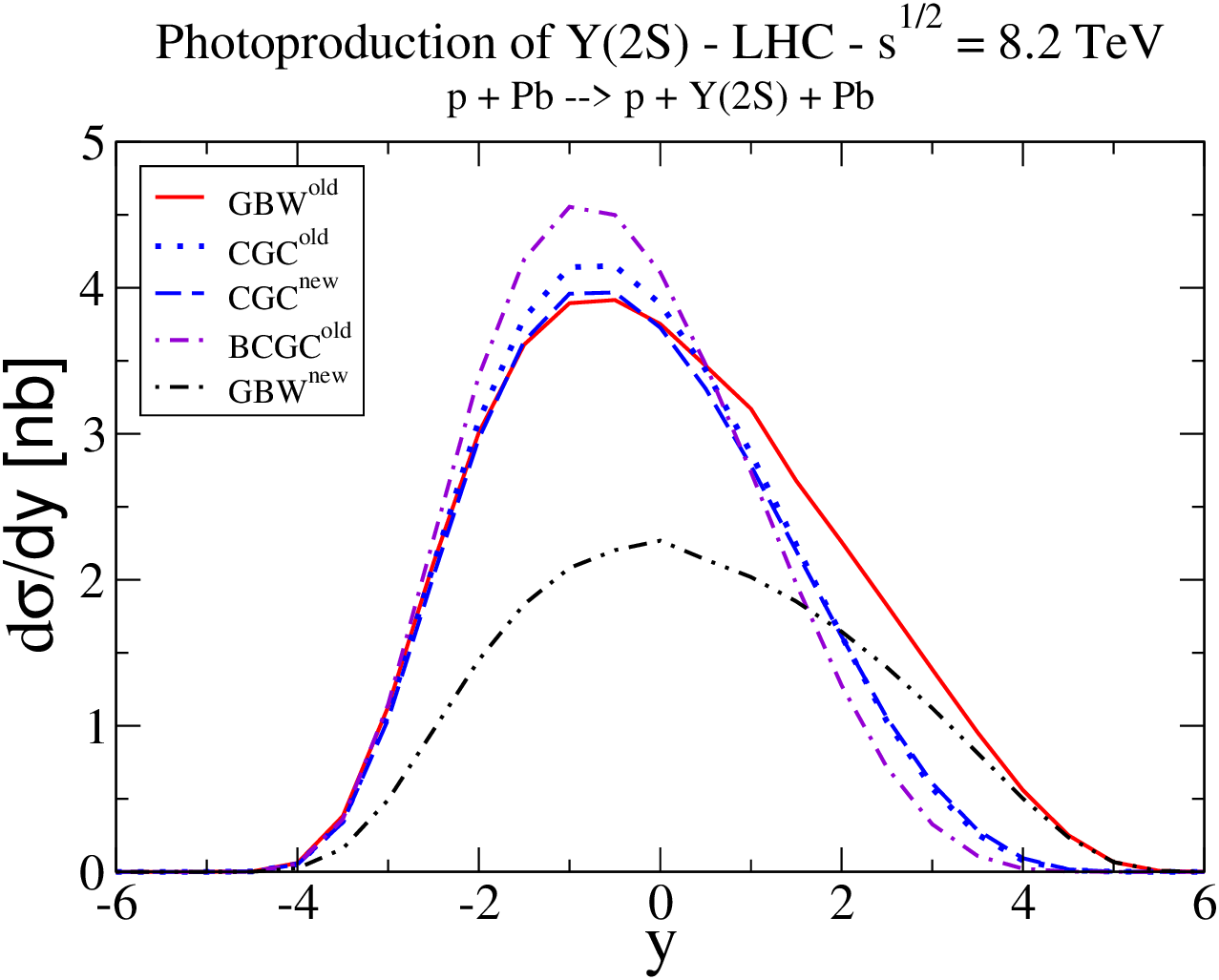} 
\includegraphics[scale=0.25]{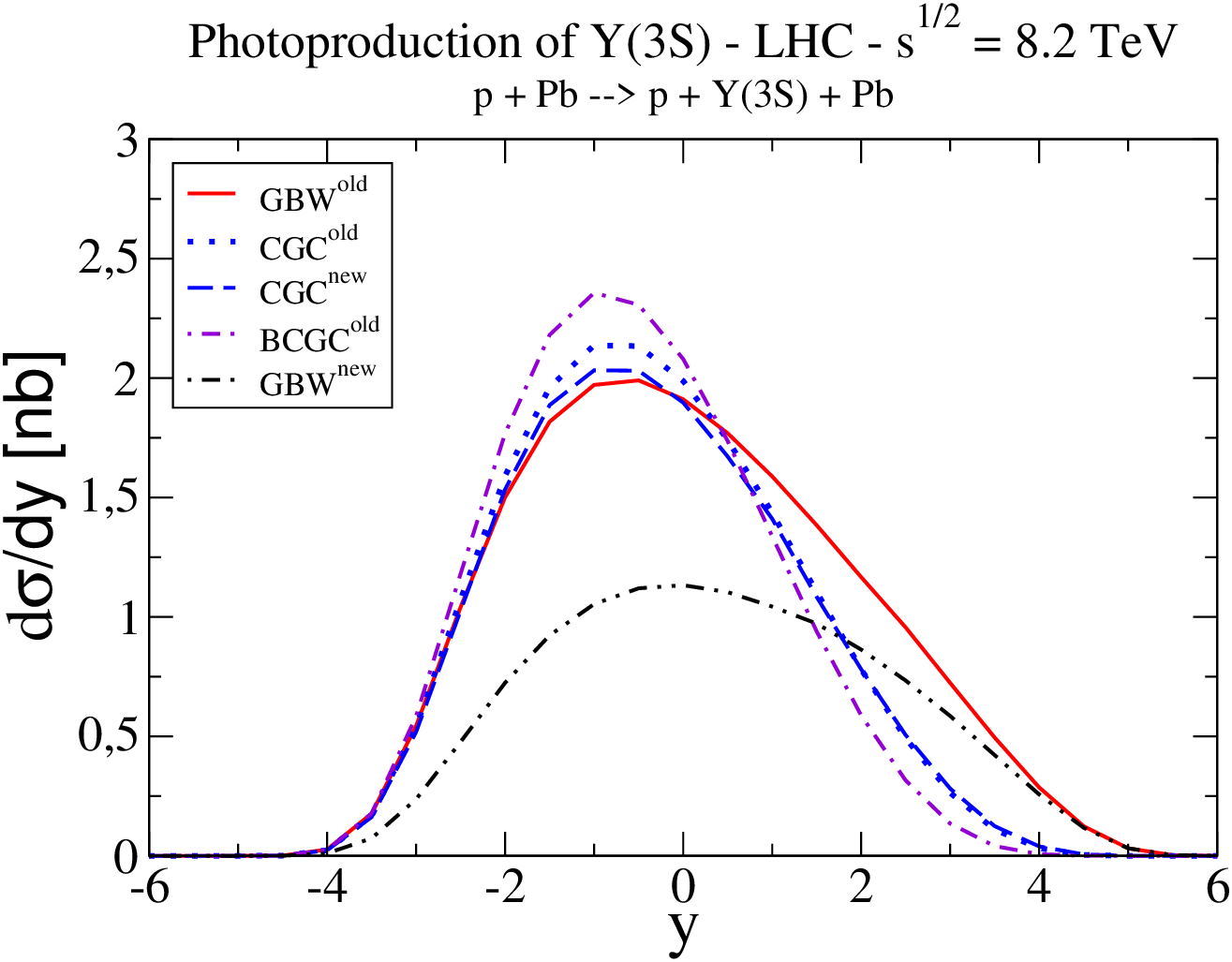} 
\caption{Rapidity distribution for the vector meson states  $\Upsilon(1S)$ (left panel), $\Upsilon(2)$ (central panel) and $\Upsilon(3S)$ (right panel)  in $pA$ collisions at $\sqrt{s}=8.2$ TeV.}
\label{Fig3}
\end{figure*}

Let us start the analysis by computing the theoretical predictions for the proton-proton case. In $pp$ collisions, the rapidity distribution of vector meson $V$  is given by 
\begin{eqnarray}   
  \frac{d\sigma}{dy}(p+p\rightarrow p+p+ V )
 & = & S^2_{gap}\left[\omega\frac{dN_{\gamma}^p(\omega)}{d\omega}\sigma(\gamma p\rightarrow V+p) \right.  \nonumber \\
&+&\left. (y\rightarrow -y)\right],
\end{eqnarray} 
where the rapidity of the produced meson is related to the photon energy by $y\simeq\text{ln}(2\omega/m_V)$. The parameter $S_{gap}^2$ quantifies the absorptive corrections\cite{antony}, and in  such a process one has the presence of large rapidity gap between the produced meson and the final state protons. In the present calculation we will take $S_{gap}^2\approx 0.8-0.9$.  Finally, the notation $(y\rightarrow -y)$ indicates the symmetry target-projectile in the $pp$ collision. In Figure $\ref{dsigdyppY1S}$, it is presented the results for photoproduction $\Upsilon(1S)$ in $pp$ collisions considering the different models presented in the last section. The relative normalization and the overall behavior on rapidity are fairly reproduced by all the models in the forward region in comparison to the experimental results from LHCb Collaboration \cite{UpsilonLHCb}. Given the present level of the experimental uncertainties it is not possible to make definitive statements about the precision of the distinct models  investigated. The theoretical uncertainty reaches a factor two considering the same wave-function and distinct dipole cross sections. On the other hand, when the integrated cross sections for $\Upsilon(1S)$ and $\Upsilon(2S)$ are considered,  a better agreement is achieved in the frontal region $2<\eta<4.5$ by the old versions of GBW and CGC models. The corresponding results for the models and the LHCb data are shown in Table \ref{sigtotalpp7tev}. For sake of completeness the production of $\psi(1S)$ and its excited state $\psi(2S)$ \cite{LHCb1,LHCb2} have been added (including also the experimental result \cite{UpsilonLHCb}), corrected for acceptance. 

\begin{figure*}[t]
\includegraphics[scale=0.25]{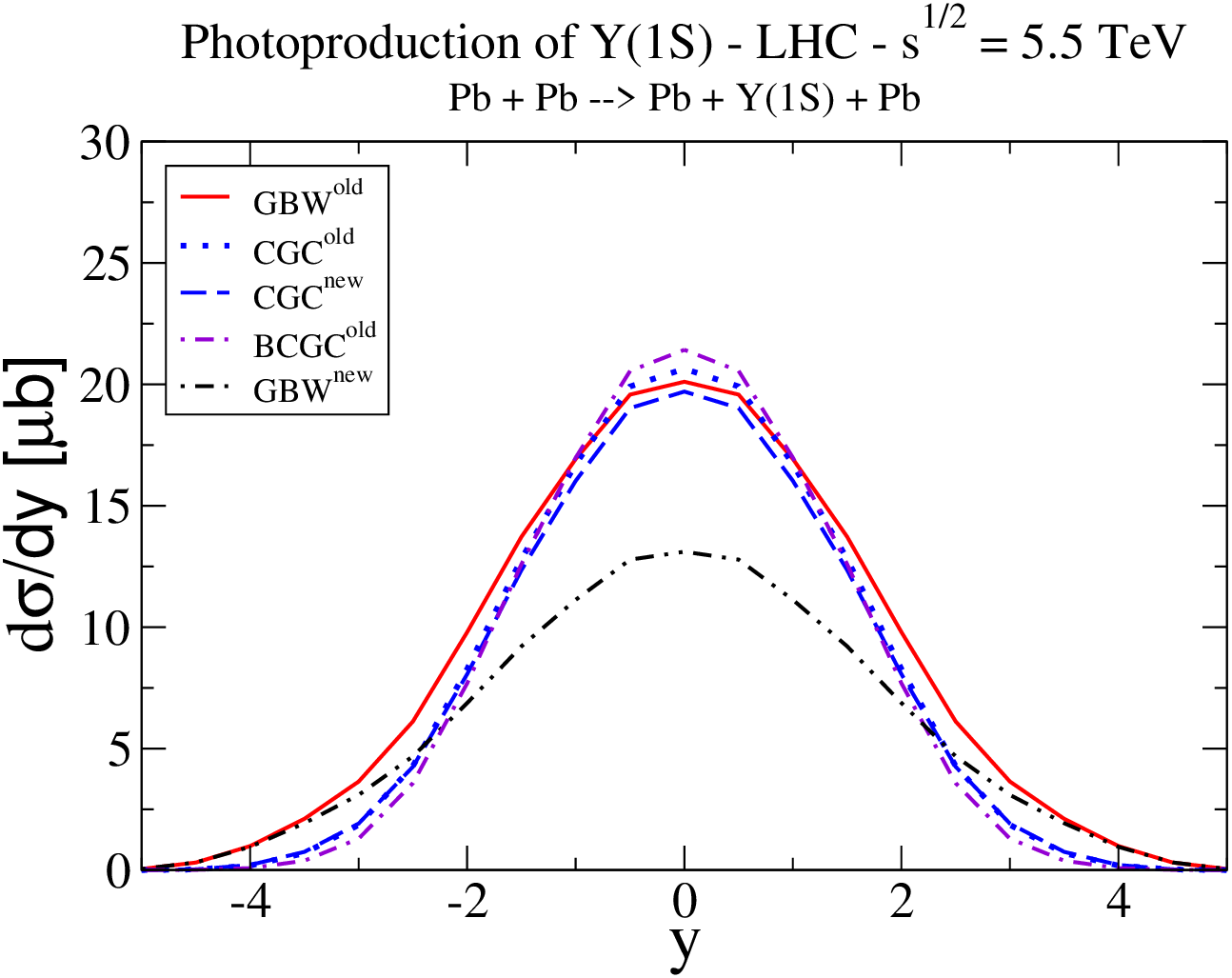} 
\includegraphics[scale=0.25]{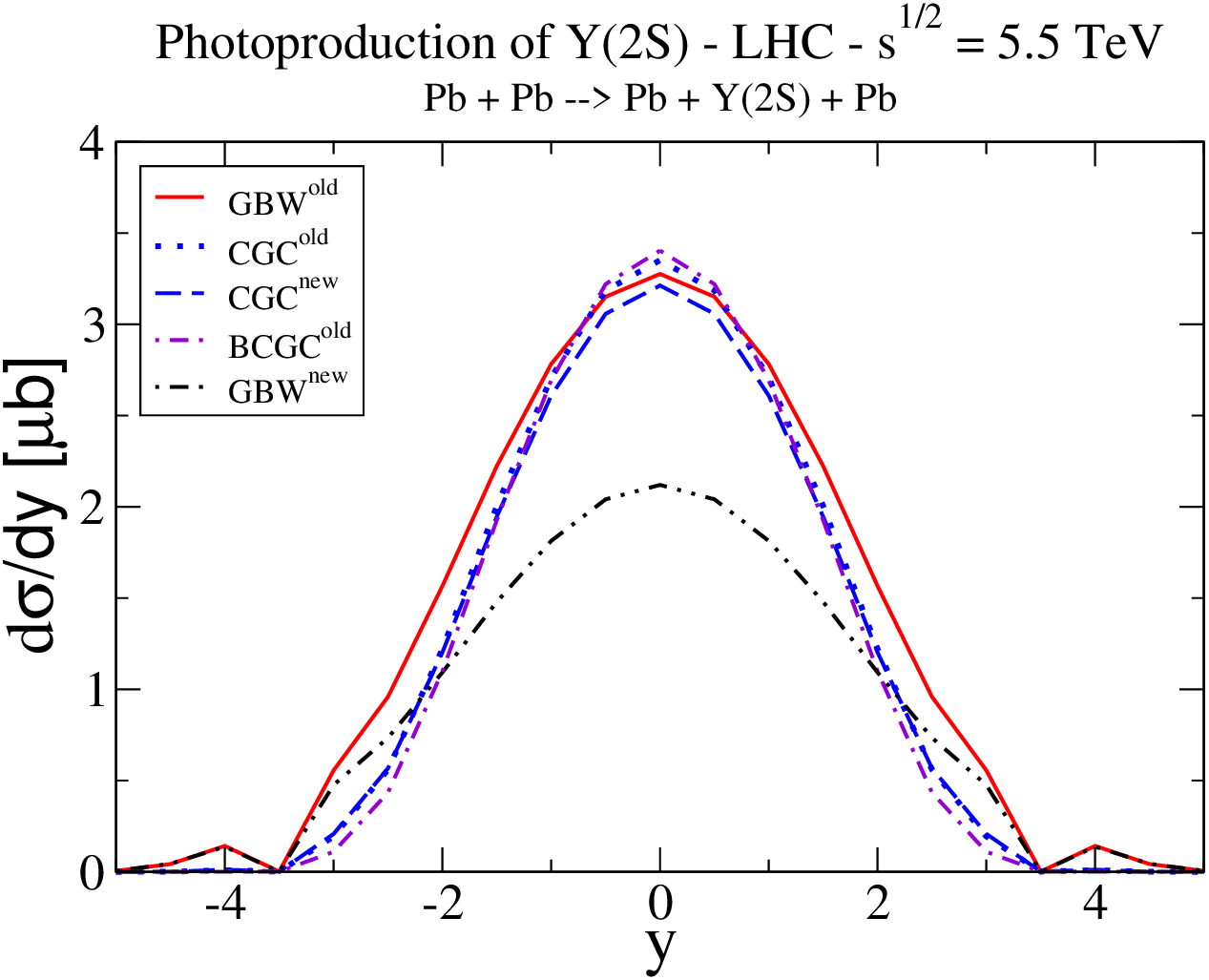} 
\includegraphics[scale=0.25]{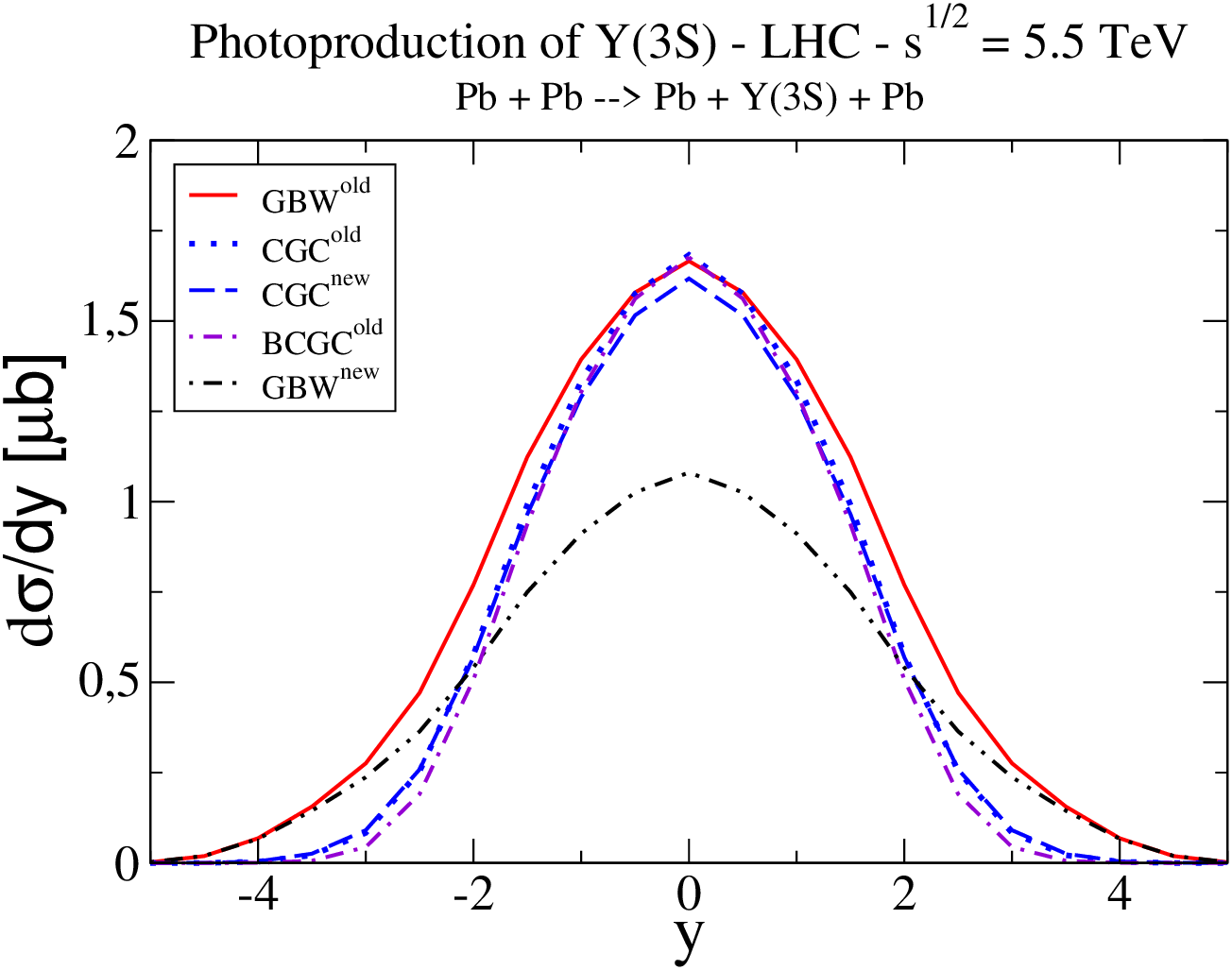} 
\caption{Rapidity distribution for the $\Upsilon(1S,2S,3S)$ states in $PbPb$ collisions at $\sqrt{s}=5.5$ TeV.}
\label{Fig4}
\end{figure*}

The calculation discussed above can be directly compared to another theoretical approaches. For instance, the present result is consistent with the NLO pQCD analysis done in Ref. \cite{Teubner}, which also compares the outcoming predictions to the LHCb data. It is verified that the theoretical uncertainty from color dipole approach is somewhat smaller than the presented there (particularly, LO and NLO pQCD predictions present large disparities). Moreover, it is worth to note that the present work updates the analysis done in Ref. \cite{Sanda2}, where the photoproduction of $\Upsilon$ states in $pp$ collisions has been done for the first time in the scope of color dipole approach, also used in \cite{Vic1} for $\Upsilon (1S)$. The focus here is to provide predictions for the next (and current) runs of LHC.

\begin{table}[t]
\scalebox{0.8}{
\begin{tabular}{ccccccc}
\hline\hline
 & \textbf{GBW} & \textbf{CGC$^{old}$} & \textbf{CGC$^{new}$} & \textbf{BCGC$^{old}$} & \textbf{GBW$^{ksx}$} & \textbf{LHCb}\\
\hline\hline      
$\psi(1s)$      & 277.60  &  213.69 &  199.58 &  154.57 &  170.81 & $291\pm 20.24$ \\
$\psi(2s)$      &  8.40   &   5.94  &   5.98  &   4.13  &   4.39  & $6.5\pm 0.98$  \\
\hline
$\Upsilon(1s)$  &  25.05   &   20.45 &   20.02  &   19.12  &  12.5  & $9.0\pm 2.7$   \\
$\Upsilon(2s)$  &  4.32    &   3.8   &   3.70   &   3.9    &   2.05 & $1.3\pm 0.85$  \\
$\Upsilon(3s)$  &  2.20    &   2.0   &   1.92   &   2.07   &   1.05 & $<\,3.4$       \\
\hline\hline
\end{tabular}}
\label{sigtotalpp7tev}
\caption{Integrated cross sections (in units of $pb$) for photoproduction of the $\psi(1S,2S)$ (corrected for acceptance) and $\Upsilon(1S,2S,3S)$ states in $pp$ collisions at $\sqrt{s}=7$ TeV compared to the LHCb data \cite{LHCb1,LHCb2,UpsilonLHCb} (errors are summed into quadrature).}
\end{table}

In Figure \ref{Fig2} predictions are done for the LHC Run 2 at $\sqrt{s}=13$ TeV in the $pp$ mode (the notation is the same as previous plot).  The general trend follows the one predicted at lower energy, including the theoretical uncertainty and overall behavior. The cross sections are quite sizeable and the relative contribution of the radial excited states compared to the lowest state follows the pattern shown at $y=0$. Namely, $\sigma [\Upsilon(2S)]/\sigma [\Upsilon(1S)]\simeq 0.17$  and $\sigma [\Upsilon(3S)]/\sigma [\Upsilon(1S)]\simeq 0.083$. There is a discrepancy at forward rapidities  when GBW-old is compared to the other models. The reason could be the typically higher saturation scale associated to that model ($Q_{sat}^2=1$ GeV$^2$ at $x = 4\times 10^{-5}$ for GBW-old, whereas it reaches one  at $x=5.5\times 10^{-7}$ for GBW-KSX). In Table \ref{sigtotalpp13tev} , the integrated cross sections are shown for the states $\Upsilon(1S,2S,3S)$, and $\psi(1S,2S)$ corrected for acceptance. Typically the cross sections are 40\% higher compared the LHC Run I at 7 TeV. There is a difference among the shapes of the distributions obtained from the dipole cross sections considered. One could therefore expect that measurements of the rapidity distributions would be able to discriminate between models. Our predictions for the $\Upsilon$ state ratios are lower that those predicted by STARlight Monte Carlo, as presented in Ref. \cite{STARlight}. The origin can be the fact that the different states are obtained from an extrapolation of HERA-DATA and using a fixed ratio for the distinct states in \cite{STARlight}. In our case, the evolution on energy is dynamically generated by parton saturation approach models and the meson wavefuntions have non-trivial behavior on the overlap function.

\begin{table}[h]
\begin{tabular}{cccccc}
\hline\hline
 & \textbf{GBW} & \textbf{CGC$^{old}$} & \textbf{CGC$^{new}$} & \textbf{BCGC$^{old}$} & \textbf{GBW$^{ksx}$} \\
\hline\hline   
$\psi(1s)$      & 997.52 &  747.75 &  696.25 &  523.3 &  598.96  \\
$\psi(2s)$      & 31.92  &    21.9 &   22.02 &  14.52 &    16.15 \\
$\Upsilon(1s)$  & 43.77  &   34.3  &   33.8  &  30.97 &    20.6  \\
$\Upsilon(2s)$  & 7.72   &   6.5   &   6.37  &  6.45  &   3.45   \\
$\Upsilon(3s)$  & 3.95   &   3.42  &   3.35  &  3.47  &   1.77   \\
\hline\hline
\end{tabular}
\label{sigtotalpp13tev}
\caption{Predictions for integrated cross sections (in units of $pb$) for photoproduction of the $\psi(1S,2S)$ and $\Upsilon(1S,2S,3S)$ states in $pp$ collisions at the LHC Run II ($\sqrt{s}=13$ TeV).}
\end{table}

\begin{figure*}[t]
\includegraphics[scale=0.28]{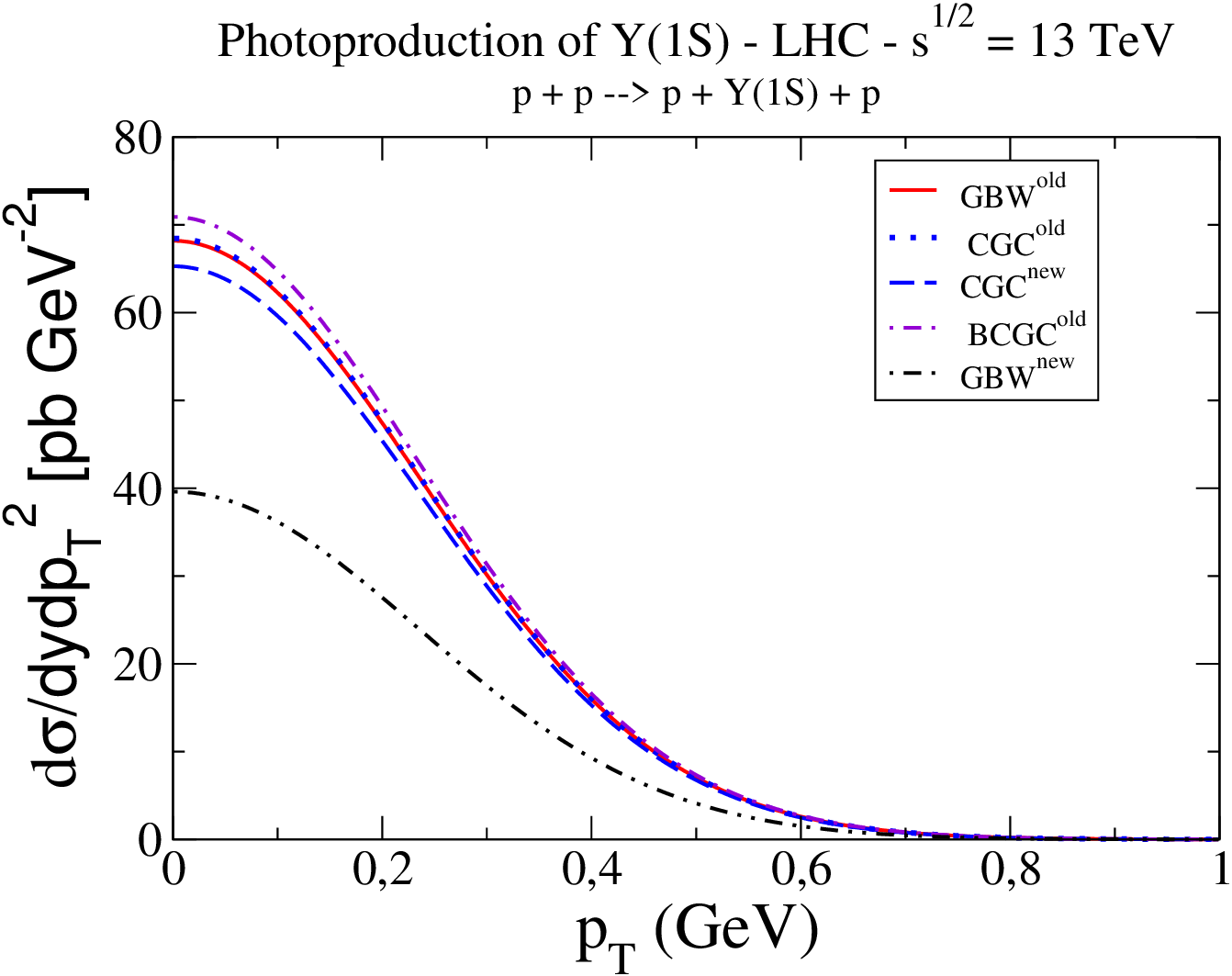} 
\includegraphics[scale=0.28]{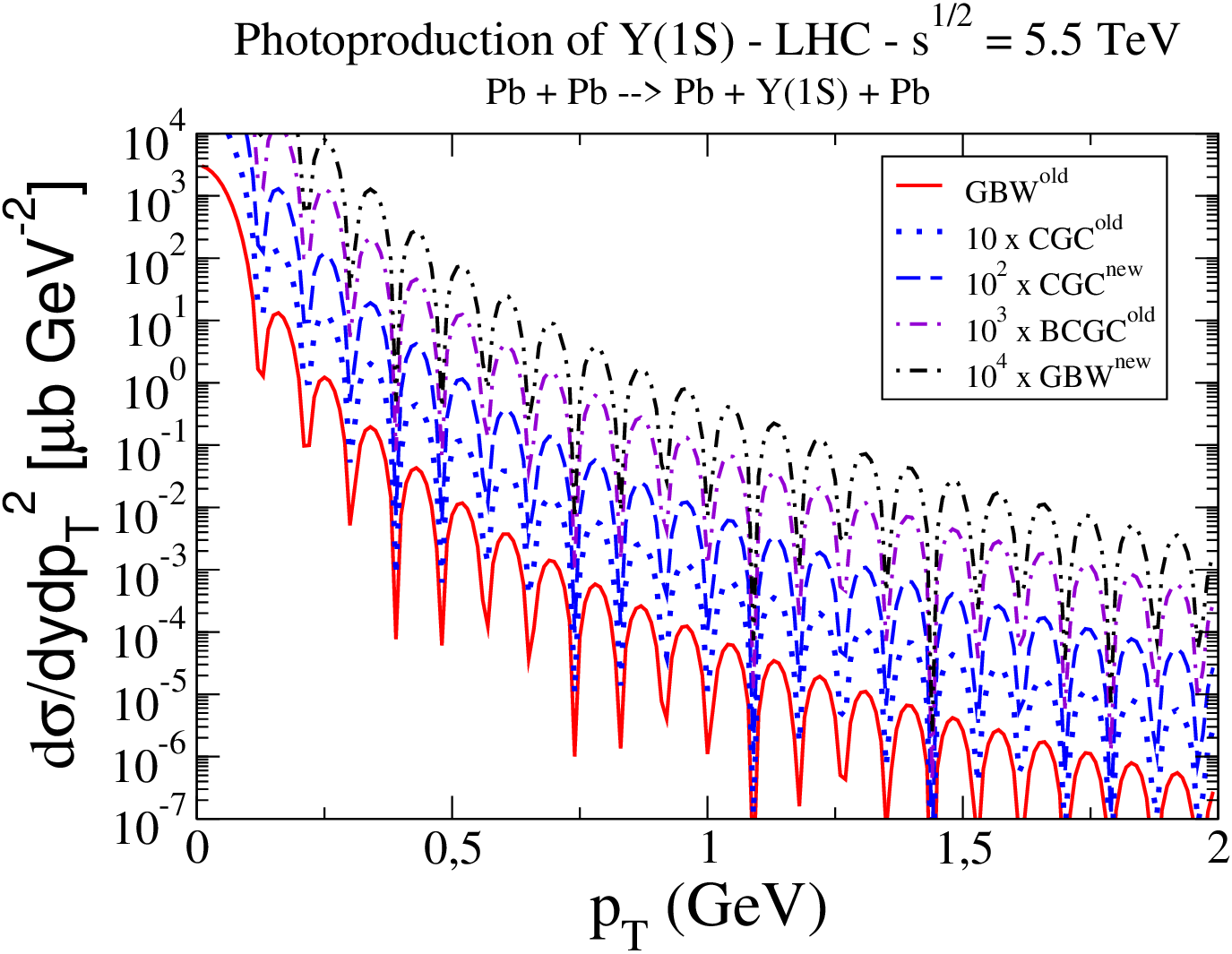} 
\caption{Transverse momentum distribution for $\Upsilon(1S)$ state in $pp$ collisions at $\sqrt{s}=13$ TeV (left panel) and $PbPb$ collisions at $\sqrt{s}=5.5$ TeV (right panel).}
\label{Fig5}
\end{figure*}

We now turn to the prediction in $pA$ ultra-peripheral collisions. In particular, in proton-lead collisions if the quarkonium rapidity, $y$, is positive in the nucleus beam direction its rapidity distribution reads as \cite{upcs}:
\begin{eqnarray}
\frac{d\sigma }{dy} (Pb+p\rightarrow Pb+p+V) & = & \frac{dN_{\gamma}^{Pb}(y)}{d\omega }\sigma_{\gamma p \rightarrow V +p}(y) \nonumber \\
& + &\frac{dN_{\gamma}^p(-y)}{d\omega }\sigma_{\gamma Pb \rightarrow V +Pb}(-y),\nonumber
\end{eqnarray}
where $\frac{dN_{\gamma}(y)}{d\omega }$ is the corresponding photon flux . The case for the inverse beam direction is straightforward. We use the Weisz\"acker-Williams method to calculate the flux of photons  from a charge $Z$ nucleus as referred in the previous section. In the numerical calculation we disregard the contribution coming from the photon flux related to the proton source. For comparison, concerning the $\psi(1S)$ and $\psi(2S)$ states at $\sqrt{s}=5.02$ TeV, the integrated cross section (in the range $ 2\leq y\leq 4.5$) is predicted to be $243.41\pm 20.37$ pb and $4.97\pm 0.48$ pb, respectively. The error includes the theoretical uncertainty related to the model of the dipole cross section. When considering the higher energy of $\sqrt{s}=8.2$ TeV the values found are $340.51\pm 36.76$ pb and $7.14\pm 0.85$ pb. Moreover, for the lowest state $\Upsilon(1S)$ one has $2.92\pm1.43$ pb at $\sqrt{s}=5.02$ TeV and $5.45\pm 2.04$ pb at $\sqrt{s}=8.2$ TeV.

In Figure \ref{Fig3} the rapidity distribution is shown for the three $\Upsilon$ states in $pA$ collisions at the energy $\sqrt{s}=8.2$ TeV. The model deviation is more intense on the very forward rapidity and the theoretical uncertainty reaches a factor two similarly to the $pp$ case. The rapidity distribution exhibits directly the influence of the $x$-dependence of the dipole cross section in the interval $0\leq y\leq 4$, and specially for rapidities around $y=3$. The shapes are similar for the lowest state and its radial excited states. Therefore, it is feasible that a consideration of $\Upsilon$ photoproduction in this rapidity interval offers potential in discriminating models of dipole cross sections. It is timely having a measurement of $\Upsilon$ production at mid-rapidity, as the corresponding cross section in the $\psi$ case it is a challenge when considering the color dipole approaches. The predictions can be directly compared to the work in Ref. \cite{AN}, where the perturbative two-gluon exchange formalism has been considered. We have not verified a second peak in the rapidity distribution as presented in \cite{AN}. The reason is that the dynamics embedded in the color dipole approach considered here corresponds to strong shadowing corrections.  The ratio $\sigma [\Upsilon(2S)]/\sigma [\Upsilon(1S)]$  and $\sigma [\Upsilon(3S)]/\sigma [\Upsilon(1S)]$ are still of same order of magnitude compared to the proton-proton case.

Finally, let us consider the $AA$ collisions.  For coherent $AA$ ultra-peripheral collisions, the rapidity distribution of vector meson $V$  is given by 
\begin{eqnarray}   
  \frac{d\sigma}{dy}(A+A\rightarrow A+A+ V )
 & = & \left[\omega\frac{dN_{\gamma}^{Pb}(\omega)}{d\omega}\sigma(\gamma A\rightarrow V+A) \right.  \nonumber \\
&+&\left. (y\rightarrow -y)\right].
\end{eqnarray} 
Figure \ref{Fig4} shows the rapidity distribution for the $\Upsilon$ states at the energy  $\sqrt{s}=5.5$ TeV (the notation is the same as previous figures). The distributions are symmetric about mid-rapidity and also similar in structure and the influence of distinct models for dipole cross section is significantly less evident than in $pA$ case. Cross section are still sizeable and the relative contribution of radial excited states compared to $\Upsilon (1S)$ is given by $\sigma [\Upsilon(2S)]/\sigma [\Upsilon(1S)]\simeq 0.15$  and $\sigma [\Upsilon(3S)]/\sigma [\Upsilon(1S)]\simeq 0.075$, which is slightly smaller than in $pp$ case. Now, the mid-rapidity region seems to show sensitivity to discriminate models of dipole cross section and corresponding nuclear effects. Compared to the STARlight Monte Carlo \cite{STARlight}, now our predictions are quite similar for the $\Upsilon (1S)$ state (we show the large theoretical uncertainty in modeling the dipole cross sections) but lower values for the $\Upsilon (2S,3S)$ states. In general, the Glauber model considered in \cite{STARlight} involves less nuclear shadowing than the Glauber-Gribov approach we have used. In Ref. \cite{AN}, the $\Upsilon$ production also was analyzed in $AA$ collisions and we verify that the theoretical uncertainty in the color dipole approach is comparable to that present in perturbative two-gluon exchange formalism. In present study only the coherent channel is considered. An important investigation would be the inclusion of ultra-peripheral collisions accompanied by photonuclear breakup. The procedure for obtaining it involves the multiplication of original photon flux by a probability factor for the desired nuclear breakup.

It was also explored the formulation to obtain the transverse momentum distribution for the exclusive photoproduction of $\Upsilon$ states. Figure \ref{Fig5} presents some samples of results concerning the $1S$ state. Left panel shows the prediction for $pp$ collisions in the energy value of the LHC Run II, $\sqrt{s}=13$ TeV. In addition, right panel presents the same distribution in $AA$ collisions at $\sqrt{s}=5.5$ TeV. For the $pp$ case, the Gaussian behavior is natural as we are imposing an exponential behavior for the $t$-dependence which is ad hoc and consistent with the experimental results from DESY-HERA on vector meson photoproduction. On the other hand, for the $AA$ case, the rich structure comes from the impact parameter dependence of nuclear dipole amplitude. Concerning the $p_T$ behavior in $pp$ collisions, it was discussed in Ref. \cite{Sanda2} that proton detectors and cuts on meson transverse momentum facilitate a measurement of $\gamma p\rightarrow \Upsilon p$ cross section at $W_{\gamma p}\approx 1$ TeV, which corresponds to the range where saturation effects are expected to be revealed.

\section{Summary}

We have considered exclusive photoproduction of $\Upsilon$ states in ultra-peripheral $pp$, $pPb$ and $PbPb$ collisions at LHC, using the color dipole approach as the underlying theoretical framework.  The rapidity and transverse momentum distributions, integrated cross section and cross section ratios for these collisions have been presented and compared with available data. The significative dependence on the dipole cross section  could be useful in discriminating models, mostly in the proton-nucleus case at forward rapidities. Our main goal is to provide predictions taking into account a theoretical framework which correctly describes  the radially excited states.

\begin{acknowledgments}
This work was  partially financed by the Brazilian funding
agency CNPq and Rio Grande do Sul funding agency FAPERGS.

\end{acknowledgments}

\end{document}